\documentclass[journal]{IEEEtran}

\usepackage{amsmath}
\usepackage{verbatim}
\usepackage{xfrac}
\usepackage{float}
\usepackage[strings]{underscore}
\usepackage[hyphens]{url}

\usepackage{cite}

\begin{document}
%
\title{A High Stability Optical Shadow Sensor with Applications for Precision Accelerometers}

\author{Steven~G.~Bramsiepe, David~Loomes, Richard~P.~Middlemiss, Douglas~J.~Paul~\IEEEmembership{Senior Member,~IEEE} and Giles~D.~Hammond%

\thanks{Manuscript received \today; revised ?}%
\thanks{S.G.Bramsiepe was supported by EPSRC under grant EP/L016753/1. The work was supported by EPSRC QuantIC grant EP/MO1326X/1.}%
\thanks{Steven~G.~Bramsiepe, Richard~P.~Middlemiss and Giles~D.~Hammond are with the University of Glasgow, SUPA School of Physics and Astronomy, Kelvin Building, University Avenue, Glasgow, G12 8UU, U.K. (Email: Giles.Hammond@glasgow.ac.uk)}%
\thanks{David~Loomes is with }%
\thanks{Douglas~J.~Paul is with the University of Glasgow, School of Engineering, Rankine Building, Oakfield Avenue, Glasgow, G12 8LT, U.K..}}%



\markboth{IEEE Sensors Journal,~Vol.~-, No.~-, Month~Year}%
{Shell \MakeLowercase{\textit{et al.}}: Bare Demo of IEEEtran.cls for IEEE Journals}

\maketitle

\begin{abstract}
Gravimeters are devices which measure changes in the value of the gravitational acceleration, \textit{g}. This information is used to infer changes in density under the ground allowing the detection of subsurface voids; mineral, oil and gas reserves; and even the detection of the precursors of volcanic eruptions. A micro-electro mechanical system (MEMS) gravimeter has been fabricated completely in silicon allowing the possibility of cost e-effective, lightweight and small gravimeters. To obtain a measurement of gravity, a highly stable displacement measurement of the MEMS is required. This requires the development of a portable electronics system that has a displacement sensitivity of $\leq 2.5$ nm over a period of a day or more. The portable electronics system presented here has a displacement sensitivity $\leq 10$ nm$/\sqrt{\textrm{Hz}}$ ($\leq 0.6$ nm at $1000$ s). The battery power system used a modulated LED for measurements and required temperature control of the system to $\pm$ 2 mK, monitoring of the tilt to $\pm$ 2 $\mu$radians, the storage of measured data and the transmission of the data to an external server.

\end{abstract}


%
\IEEEpeerreviewmaketitle

\section{Introduction}
%
%
%
%
\IEEEPARstart{G}{ravimeters} are devices used to measure the local acceleration of gravity, \textit{g}. They can measure the absolute value of \textit{g}, an absolute gravimeter, or the changes in \textit{g}, a relative gravimeter \cite{Wolfgang1989}. They have many practical uses such as in industry for oil, gas and mineral surveys \cite{Barnes2012}. Gravimeters can also be used in environmental monitoring of volcanoes, giving advanced warning before eruptions \cite{Battaglia2008}\cite{Carbone2017} and to measure density changes in water, which can be used for climate models \cite{Chambers2004}. Current relative systems cost in excess of $\pounds 60000$, weigh more than $5$ kg and are cumbersome to carry [Scintrex CG5 and CG6\cite{ScintrexCG5}]. Recently a Micro-electro Mechanical System (MEMS) gravimeter has been fabricated and tested demonstrating promise for small, low-cost and lightweight gravimeters \cite{Middlemiss2016}\cite{Campsie}. 


The MEMS, made of silicon, consists of a central proof mass connected by three anti flexures \cite{Middlemiss2016}\cite{Middlemiss2016a} and can be thought of simply as a mass on a spring, constrained in one dimension. Below the resonant frequency of the device, there is a constant relationship between the acceleration and the displacement of the central proof mass. This change in displacement, if measured, can therefore allow the change in acceleration to be calculated. An important point to note is that this system is a relative gravimeter that can determine changes in the gravitational acceleration, \textit{g}. i.e. it can determine that \textit{g} has changed by \textit{$\delta$g} either as a function of time or position. To measure a useful \textit{$\delta$g} of around $40$ $\mu$Gal/$\sqrt{\textrm{Hz}}$ \footnote{$1$ Gal $\equiv$ $10$ mm s$^{-2}$ $\approx$ $1$ mg}, a displacement sensitivity of $\leq 2.5$ nm/$\sqrt{\textrm{Hz}}$ for a $2$ Hz resonator is required. 

The displacement is measured using the lock-in technique\cite{Blair2001} with a shadow sensor (seen in figure \ref{fig-shadowsensor}). The shadow sensor operates by illuminating the proof mass using an LED, causing a shadow to fall on the photodiodes as the position of the proof mass changes. Lock-in amplifiers are commonly used to remove \sfrac{1}{f} noise by modulating the signal at a higher frequency before demodulating to obtain the noise performance around the modulated frequency and removing the $\sfrac{1}{f}$ noise with a low pass filter\cite{Blair2001}.

To obtain the necessary displacement sensitivity of the MEMS proof mass, a large suite of electronics is required to: modulate and demodulate the LED/signal, measure and control the temperatures via use of digital to analogue converters (DACs), convert the photocurrent to a voltage (IV converter), convert the analogue signals to digital signals and to perform filtering. The reason for implementing the temperature control is that the system is sensitive to temperature changes, both via the LED and the MEMS itself. Temperature changes affect the Young's modulus of the silicon and the coefficient
of expansion of silicon results in a 2.6 $\mu$m displacement per 1 K temperature change. Therefore the temperatures of the proof mass and flectures were required to be controlled to within $\pm 2$ mK to obtain the necessary sensitivity from the shadow sensor. 

To utilise the small, lightweight nature of the MEMS, the electronics surrounding the device had to be miniaturised. The objective was to reduce a $19$" electronics rack down to a battery powered, portable size whilst still keeping or improving the current performance and functionality. There were no off-the-shelf electronics that had all the requirements necessary, so a custom board had to be designed. The electronics board would utilise a micro-processor to communicate with each of the components and compute digital filters.

 


It was noted that as the system was miniaturised, a small and cost effective lock-in amplifier would have to be designed and created. It was observed from \cite{Sonnaillon2005}\cite{Bengtsson2012}\cite{Carrato1989}\cite{Leis2012} that digital lock in amplifiers could have the necessary performance for the gravimeter system and could even be implemented in low cost micro-controllers \cite{Dorrington2002}\cite{Aguirre2011} such as the dsPIC33E \cite{MicrochipdsPIC33}\cite{Wenn2007}.

Here, a micro-controller would digitally filter and demodulate the signal from the photodiodes unlike the previous system\cite{Middlemiss2016}, which used a LIA-MV(D)-200-L\cite{LASERLIAMVD} to demodulate the signal. The micro-controller then performs a digital low pass filter and decimates the data to obtain the performance. Further thought behind designing a custom board is that many of the individual pieces of electronic equipment required all cost in excess of $\pounds 1000$ each. This would not be in keeping with the main objective of portability and affordability of the complete system.


Since the MEMS is measuring the force of gravity exerted on a mass on a spring system, tilt becomes an important variable to measure. If the angle to the force of gravity changes, the device is no longer parallel to \textit{g} and instead $g\cos{\theta}$ would be measured where $\theta$ is the angle to the vertical. Therefore it was important to design into the electronics board the ability to measure the tilt of the sensor.

Here we demonstrate a micro-controller based electronics board capable of not just the displacement sensitivities required in measuring changes in \textit{g} of $\leq 40$ $\mu$Gal/$\sqrt{\textrm{Hz}}$, which, was successfully taken into the field[CITEFIELD] for gravimetry measurements, but a high stability optical displacement sensor complete with electronic readout and control that could be used for other precision sensing applications.


\section{Electronics Board}
\begin{center}
	\begin{figure}
	
	\includegraphics[width=0.45\textwidth]{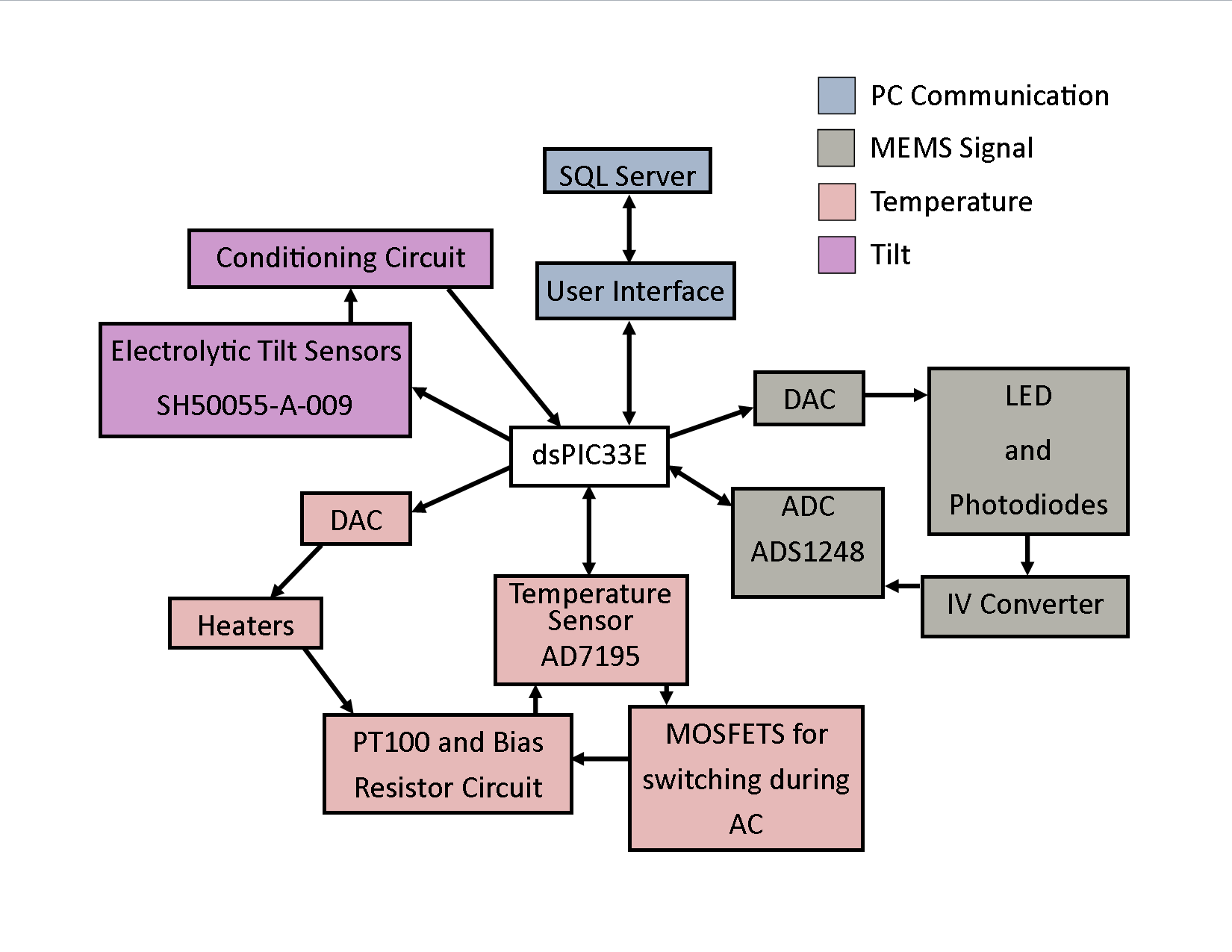}

	\caption{A block diagram outlining the functionality of the custom electronics board used to make low noise displacement measurements of a MEMS gravimeter. The block diagram is comprised of four discrete sections and are the processes involved in: communicating with the PC, measurements of the MEMS, measurements of temperature and its control and measurements of tilt.}
	\label{fig-dsPICBlock}
	\end{figure}
\end{center}

The vision of a low cost, lightweight and small MEMS gravimeter required the design and testing of a custom electronics board. This was due to no single piece of electronics combining all of the necessary functionality required to measure the small changes in gravity being commercially available. The board required the ability to: source a stable modulated drive for an LED to be used in the shadow sensor; convert the low currents from a split photodiode setup into a usable digital signal via use of a transimpedance amplifier and analogue to digital converter; measure and control temperatures via the use of a ratiometric resistance measurement and a proportional, integral and derivative (PID) controller; monitor tilt allowing regression if necessary; compute digital filters including finite impulse response (FIR) filters and decimations. The board would also require the data to be accessible in real time and allow settings to be changed also in real time. Figure \ref{fig-dsPICBlock} shows a block diagram of the necessary functions of the electronics system. 

The electronics board would be unified by a Microchip dsPIC33E micro-processor. This dsPIC could not just communicate with the electronics on the electronics board but communicate with a custom user interface (UI) on a laptop/desktop using a SQL Server framework database. The micro-controller would check during each loop for a valid connection to a computer and database and send this database several series of data packets containing information on both the analogue and digital inputs and outputs. The user interface then communicates with the database to collect the latest values and then updates it with the settings chosen on the interface. This custom UI also had the ability to plot graphs in real time and take logs for future use. It was, however, limited to refreshing values at a rate of $0.5 \rightarrow 30$ Hz due to limitations in communicating with the database. This $30$ Hz is a soft limit, however, and could probably be exceeded, it was found to be unnecessary since the decimation of the data reduces the data rate below this level.

\section{Shadow Sensor}

\begin{center}
	\begin{figure}

	\includegraphics[width=0.5\textwidth]{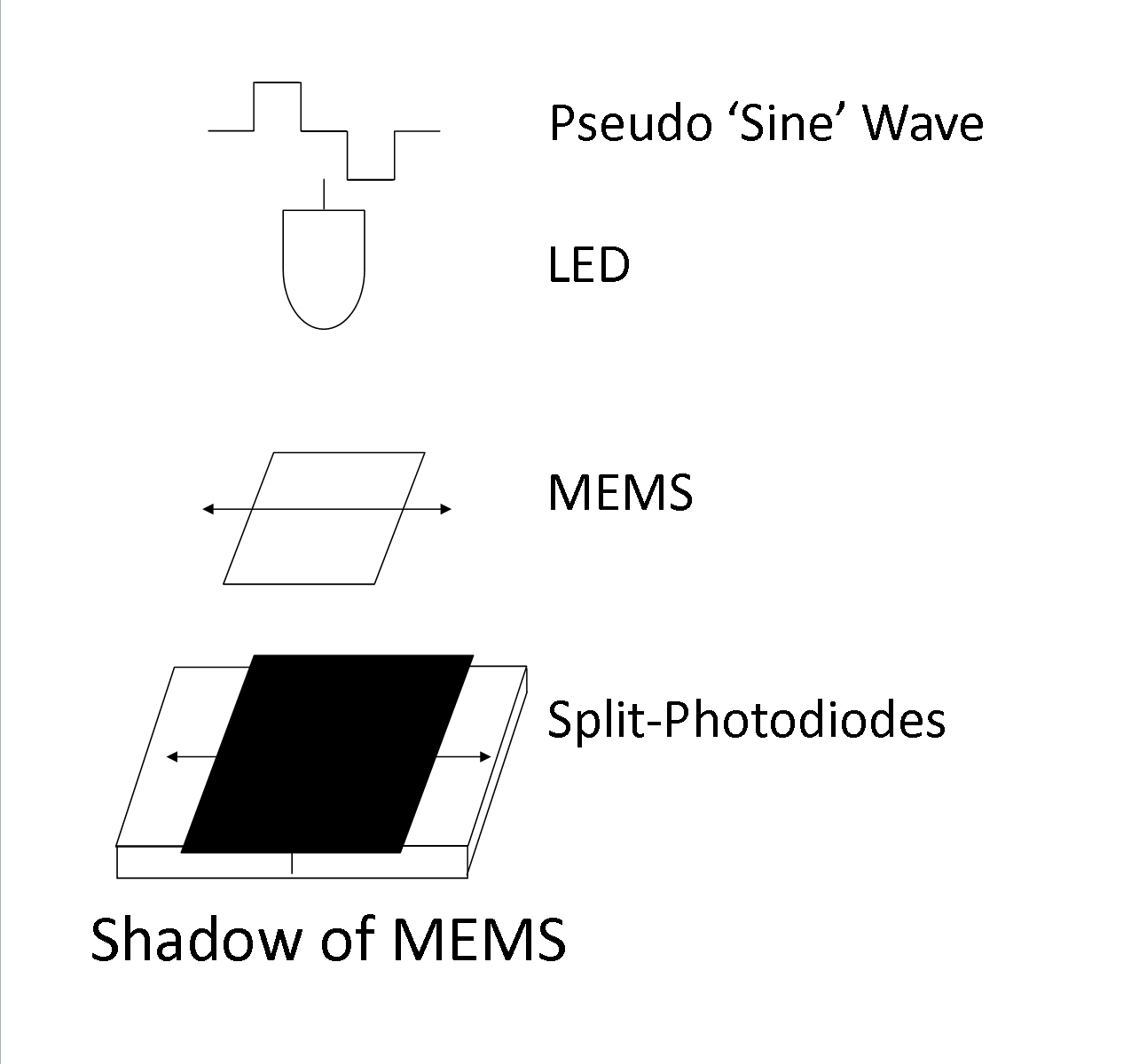}

	\caption{Illustration of the Shadow Sensor Technique. LED illuminates a moving MEMS that casts a shadow over a split photodiode that takes the difference in intensities of the two photodiodes. This allows large gains during amplification.}
	\label{fig-shadowsensor}
	\end{figure}
\end{center}

\begin{center}
	\begin{figure}

	\includegraphics[width=0.5\textwidth]{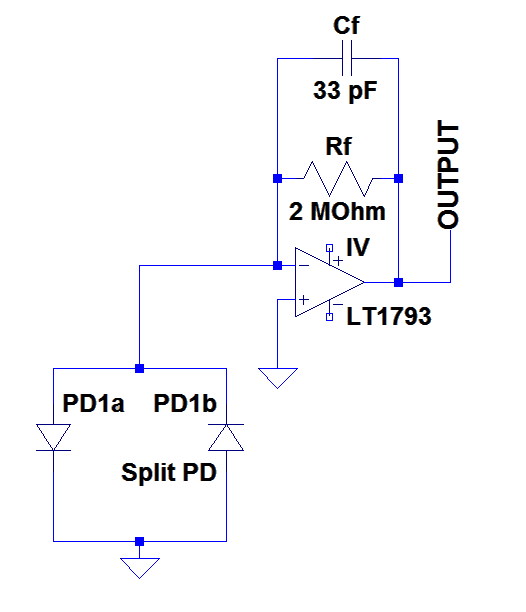}

	\caption{Circuit diagram of a split photodiode that takes the difference of the currents from each diode. This allows amplification using a current to voltage (IV) converter with a gain of $2 \times 10^{6}$.}
	\label{fig-splitphotodiode}
	\end{figure}
\end{center}

\begin{center}
	\begin{figure}

	\includegraphics[width=0.5\textwidth]{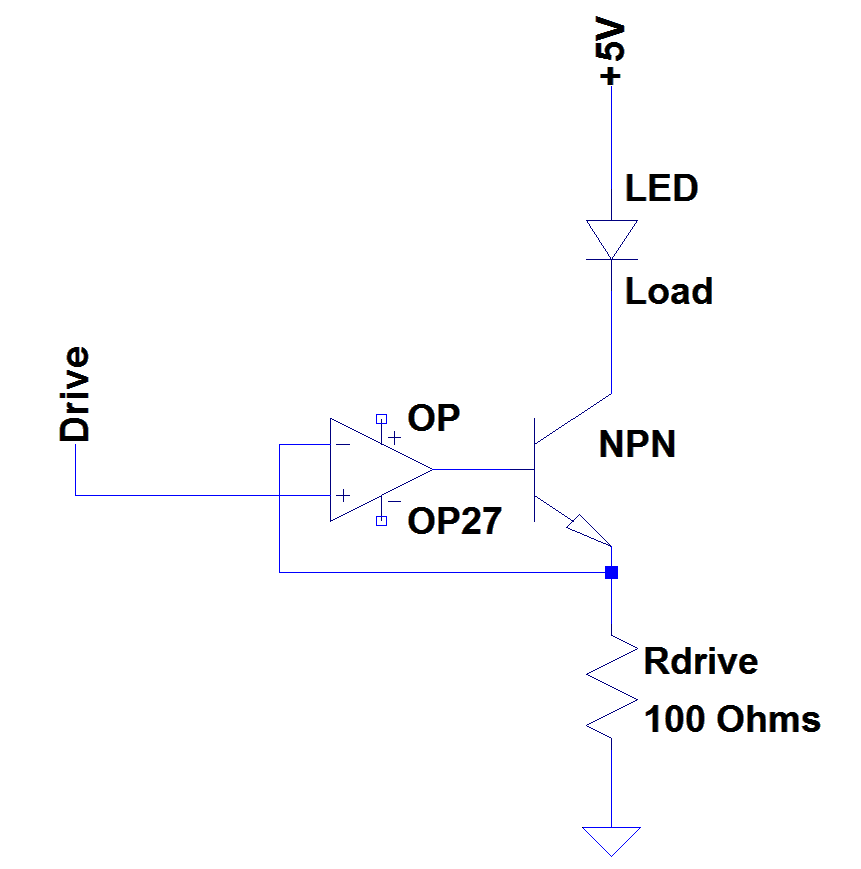}

	\caption{Circuit diagram for a stable current source. The operational amplifiers output changes the conductance of the transistor enough to ensure there is a voltage at the high side of the drive resistor equal to the one being applied as the drive. Current is equal to the voltage applied over the drive resistor over its resistance, i.e. $i = \frac{V_{drive}}{R_{drive}}$}
	\label{fig-currentsource}
	\end{figure}
\end{center}

The displacement of the proof mass is measured using a shadow sensor combined with a lock-in amplifier. In the shadow sensor, the MEMS proof mass is illuminated with an LED, creating a shadow on the photodiodes (as seen in figure \ref{fig-shadowsensor}). This photo-current is then converted to a voltage and sampled via an analogue to digital converter (ADC). Several forms of shadow sensor exist. Here, only two photodiodes in a differential setup\cite{Middlemiss2016} are used (as seen in figure \ref{fig-splitphotodiode}), however, there also exists a version using a quadrant arrangement of photodiodes\cite{Zoellner2013} but due to the proof mass being constrained in one dimension, the simpler case is sufficient. The photodiodes are operated in a photovoltaic mode to remove the current noise. The electronics convert the photocurrent form the photodiodes to a voltage using a current to voltage converter which contains a low pass filter (RC network). This low pass was designed to have a cut off frequency at $\approx 2300$ Hz even though the sampling frequency was much lower than the ADCs used to sample the signal ($640$ Hz). This was due to the nature of the lock-in setup which has only four points per cycle of the sine wave (seen at top of figure \ref{fig-shadowsensor}). If the cut off frequency was lower, rounding was observed in the output due to the higher frequency components being attenuated, not too dissimilar to a square wave. After the current is converted to a voltage, $1.5$ V must be summed onto the signal to allow for a fully differential measurement (seen on page 29 of the datasheet\cite{TIADS1248}). The summation is done using a non-inverting summing amplifier utilising a TL071. This scheme was used as due to the way the electronics board was originally designed. The ADS1248 would not be able to measure anything when an input is below ground, therefore, having a common mode input half way between the reference voltage and ground would allow for psuedo negative values (negative relative to the negative input but still positive relative to ground). The driver for the LED is derived from a stable current source circuit presented in figure \ref{fig-currentsource}. The current source ensures a current equal to V$_{\textrm{drive}}$/R$_{\textrm{drive}}$ passes through the LED. The circuit that does this can be observed in figure \ref{fig-currentsource}. The operational amplifier output changes the conductance of the NPN transistor so that the voltage above R$_{\textrm{drive}}$ is matched to the inverting terminal of the operational amplifier. This means that the current passing though the LED has to be equal to V$_{\textrm{drive}}$/R$_{\textrm{drive}}$. In an ideal case where the proof mass is perfectly centered on the shadow sensor, zero signal would be the output and, any changes in the LED would be cancelled as the variations are common in both the photodiodes. This, however, cannot be realised and thus there will always be sensitivities to optical variations due to temperature and changes in the drive current (perhaps due to R$_{\textrm{drive}}$ or V$_{\textrm{drive}}$ changing) but this does not stop the sensor from reaching the required performance. An advantage of using a differential setup (also known as a split photodiode) is that large gains can be realised due to the currents having been subtracted before the transimpedance amplifier stage. If, instead, a setup required individual amplification of each photodiode before a differential amplifier stage, then the operational amplifiers used are likely to saturate before any subtractions can be carried out. Here we obtain a gain between $(1 \rightarrow 10) \times 10^{6}$, obtaining a calibration of our signal in Volts of $\approx 20 \times 10^{3}$ V/m for a drive current of $\approx 11$ mA, i.e. we would see a change in signal of $\approx 20$ mV for a 1 $\mu$m displacement of the proof mass.

There are several forms of noise that contribute to the noise in the measured signal (of which Johnson-Nyquist noise and shot noise are examples). An estimation of the shot noise in the system can be obtained by simply covering a single photodiode. The resulting current in this setup is approximately $I = 0.4$ mA from each of the diodes. This can then be combined with equation \ref{eq:shotnoise} to obtain a shot noise of $11$ pA/$\sqrt{\textrm{Hz}}$ (or $0.5$ nA when multiplying with a bandwidth of $\Delta f = 2300$ Hz). 

\begin{equation}
\label{eq:shotnoise}
	\sigma_i = \sqrt{2 \times q \times I}
\end{equation}

\noindent where $q$ is the fundamental charge of an electron equal to $\approx 1.6 \times 10^{-19}$ C. The shot noise from each of the photodiodes is incoherent and therefore should be added together as such. This gives an overall shot noise of $\sigma_{iT} = \sqrt{2} \times 0.5 \times 10^{-9} \approx 0.7$ nA. This noise is then amplified into voltage noise through the current-to-voltage amplifier presented in figure \ref{fig-splitphotodiode}. When using a feedback loop with a resistance of $2$ M$\Omega$, the current noise becomes a voltage noise, $\sigma_v = \sigma_iT R_f = 0.7 \times 10^{-9} \times 2 \times 10^{6} = 1.4$ mV $\equiv 70$ nm. This means 70 nm of noise is on the input of the ADC across a bandwidth of $2300$ Hz. The ADC, however, only samples at 640 Hz resulting in aliased noise. An input noise of $1.4$ mV$_{rms}$ can be inserted into a simulated version of the digital filters used by the micro-controller to obtain a noise estimate of $18$ $\mu$V$_{rms}$ $\equiv 0.9$ nm$_{rms}$ (after four stages of decimation). Although this is lower than the observed level it does demonstrate that the system is close to the shot noise limit (a factor of $\approx 5$ lower).

Thermal noise, also known as Johnson-Nyquist noise, originates from random thermal motion of carriers and can be calculated using equation \ref{eq:thermalnoise}\cite{SpringerElectronicsNoise}. 

\begin{equation}
\label{eq:thermalnoise}
	\nu_n = \sqrt{4 k_B T R \Delta f}
\end{equation}
\noindent Where $k_B$ is Boltzmann's constant and $T$ is temperature $\approx 300$ K. From figure \ref{fig-splitphotodiode} it can be observed that a feedback resistor of $R = 2$ M$\Omega$ is used, with a bandwidth of $\Delta f = 2300$ Hz, so a thermal noise of $8.7$ $\mu$V$_{rms}$ can be calculated. This value is several orders of magnitude smaller than the current shot noise and can be ignored as it is not limiting the performance.

The ADS1248 also contains noise instrinsic to the inputs. These values are stated on page 21 of the datasheet \cite{TIADS1248}. For a sampling frequency of $640$ Hz, the measurement would have $15.4$ effective number of bits (peak to peak), or, $8.6$ bits of noise relative to full scale. $8.6$ bits of the $3$ V is the equivalent of $69.4$ $\mu$V of noise of the input ($\equiv 34.7$ nm). Simulating the effect of this noise through the lock-in would only result in $3 \rightarrow 4$ $\mu$V $\equiv 0.15 \rightarrow 0.20$ nm. This is an order of magnitude lower than the short term noise that is observed. Another source of noise is the circuit noise. This noise was measured to be $100$ nm peak to peak before filtering, which results in $\approx 2$ nm post filtering. Thus we expect that the system is currently limited by a combination of shot noise from the optical sensor and electronics noise from the signal conditioning amplifiers which precede the inputs of the ADS1248. Our noise model matches up well to what is seen at a PGA of $1$ in figure \ref{fig-lockin_noise}.

%
%
%
%
\section{Lock-in Amplifier (LIA)}

Lock-in amplification is a technique that can remove low frequency (1/f) noise from a measurement. This reduction is achieved by modulating signals at a higher frequency than the experiment, $f_{M}$ and then demodulating afterwards. This approach obtains noise around the frequency of modulation with the bandwidth chosen by the low pass filter used after the demodulation. 

\subsection{Digital Lock-in Amplifiers}

The electronics board creates a modulated drive signal in a unique way relative to most other systems. The DAC (TLV5616\cite{TITLV5616}) that is used for the drive signal is only changed when a sample is recorded by the ADS1248\cite{TIADS1248}. Once the micro-controller receives an updated value from the ADS1248, the firmware puts the ADS1248 to sleep and then changes the output of the DAC. After a pre-programmable number of clock cycles, the micro-controller then awakens the ADS1248 and starts it converting a sample. This process is then repeated. The pre-programmable delay is used to ensure that the ADS1248 samples once the signal has settled to a new value. The modulated signal is in the form of a `pseudo-sine wave' -- a sine wave with only 4 points per cycle, 0, 1, 0, -1 (figure \ref{fig-shadowsensor}). Using 4 points per cycle has an advantage because the ADS1248 is limited in speed, so the fewer points per cycle of sine wave, the higher the modulation frequency. This signal still contains the necessary frequency information for demodulating to a DC level and was chosen over a conventional square wave as it gives information on both the in-phase and out of phase components of the signal. Introducing a delay resulted in a modulation frequency of around $110 \textrm{Hz} \rightarrow 150 \textrm{Hz}$, depending on the programmed delay. The sampling frequency  was chosen to be $640$ Hz with a $\sim 400$ $\mu$s delay. Even though the ADS1248 can sample up to $2$ kHz, the sampling rate was chosen to be $640$ Hz since the input referred noise increases with modulation frequency (as seen on page 21 of its datasheet\cite{TIADS1248}). Although this increase in frequency would also increase the modulation frequency, it would not give better noise relative to the increased input noise. 

Once the signal is demodulated, the micro-controller filters the data using a finite impulse response (FIR) filter\cite{DKLindnerSAS}. This takes advantage of the dsPIC's specialised digital signal processing (DSP) engine that can multiply two data registers, add the result to an accumulator, fetch the next data words to the data registers and update these data registers in two clock cycles\cite{MicrochipdsPICEngine}. This engine block allows many different variables each to have their own decimation stage, allowing variables to be chosen for filtering and decimation, without slowing down the performance of the dsPIC. The filter used for decimating is a custom low pass with a cut off frequency at \sfrac{1}{8}$^{\textrm{th}}$ of the sampling frequency, allowing a decimation factor rate of 4:1 without running into aliasing\cite{DKLindnerSAS}. 

\begin{center}
	\begin{figure}
	
	\includegraphics[width=0.45\textwidth]{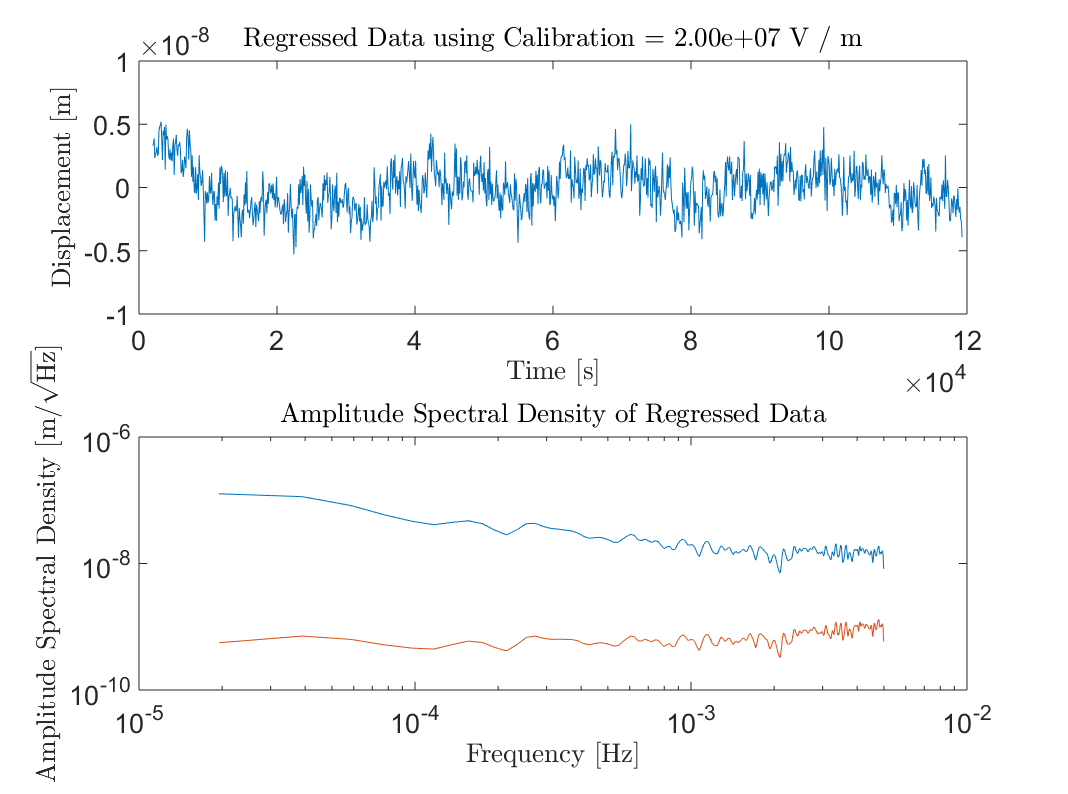}

	\caption{Output from the 4th decimation step of the lock-in signal, measuring the position of the MEMS over the course of $\approx 110$ ks. Top graph shows the time domain signal in m of which as been down-sampled to allow a regression. This regression looks at how correlated the signal and other measured variables are to obtain a fit which is subtracted from the original signal. The bottom graph shows the amplitude spectra density in m/$\sqrt{Hz}$ as well as the amplitude spectral density multiplied by the square root of the frequency giving the noise when averaging to that frequency.}
	\label{fig-lockin}
	\end{figure}
\end{center}

\begin{center}
	\begin{figure}
		\includegraphics[width=0.45\textwidth]{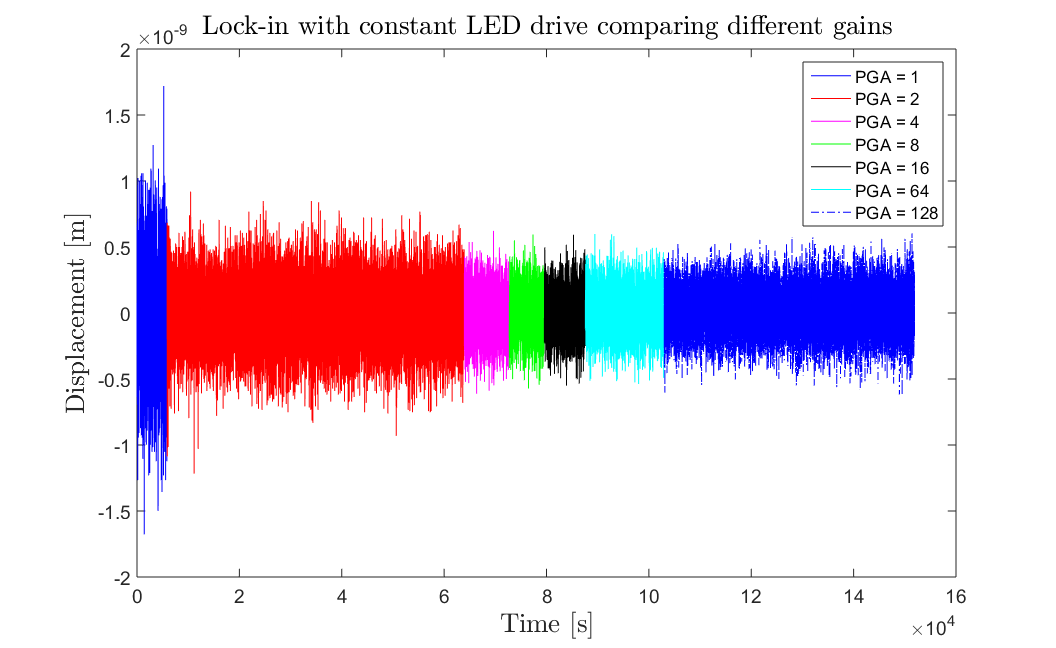}
		
		\caption{Output from the 4th decimation step of the lock-in signal without a MEMS present and no modulation on the LED. It shows initially a decreasing noise as the gain on the ADC increases. Should be noted that the peak to peak noise is lower than that when modulating the LED implying the system is limited by the LED.}
		\label{fig-lockin_noise}
	\end{figure}
\end{center}

Figure \ref{fig-lockin} demonstrates the typical  rms displacement of the proof mass measured over more than a day as well as the amplitude spectral density. The noise floor sits at $\sim 0.6$ nm at $1000$ s. Averaging to $10000$ s only decreases the noise by a factor of $2$ due to the \sfrac{1}{f} noise. Figure \ref{fig-lockin_noise} shows the rms noise when the LED is not being modulated but is on at a constant output. It shows a noise peak to peak of $2$ nm with no low frequency drift (compared to $4$ nm short term noise when the LED is modulated). Comparing the two figures it can be observed that modulating the LED increases the high frequency noise and introduces low frequency drift. This would imply the system is normally limited by the LED.

\section{Ratiometric Temperature Measurement}

A common way of measuring temperatures is through the use of PT100 resistive thermometers. These resistors are platinum resistors, precision made to have a resistance of $100$ $\Omega$ at $0 ^{\circ}C$ and to then have a linear increase with temperature (here $0.385$ $\Omega/K$). This property allows a calculation of the temperature to be made once the resistance is measured as demonstrated in equation \ref{eq-Temperature}. 

\begin{equation} \label{eq-Temperature}
	T = \frac{R_{measured} - 100}{0.385}
\end{equation}

\noindent where $T$ is in $^{\circ}C$, $R_{measured}$ is in $\Omega$. 

\begin{center}
	\begin{figure}

	\includegraphics[width=0.5\textwidth]{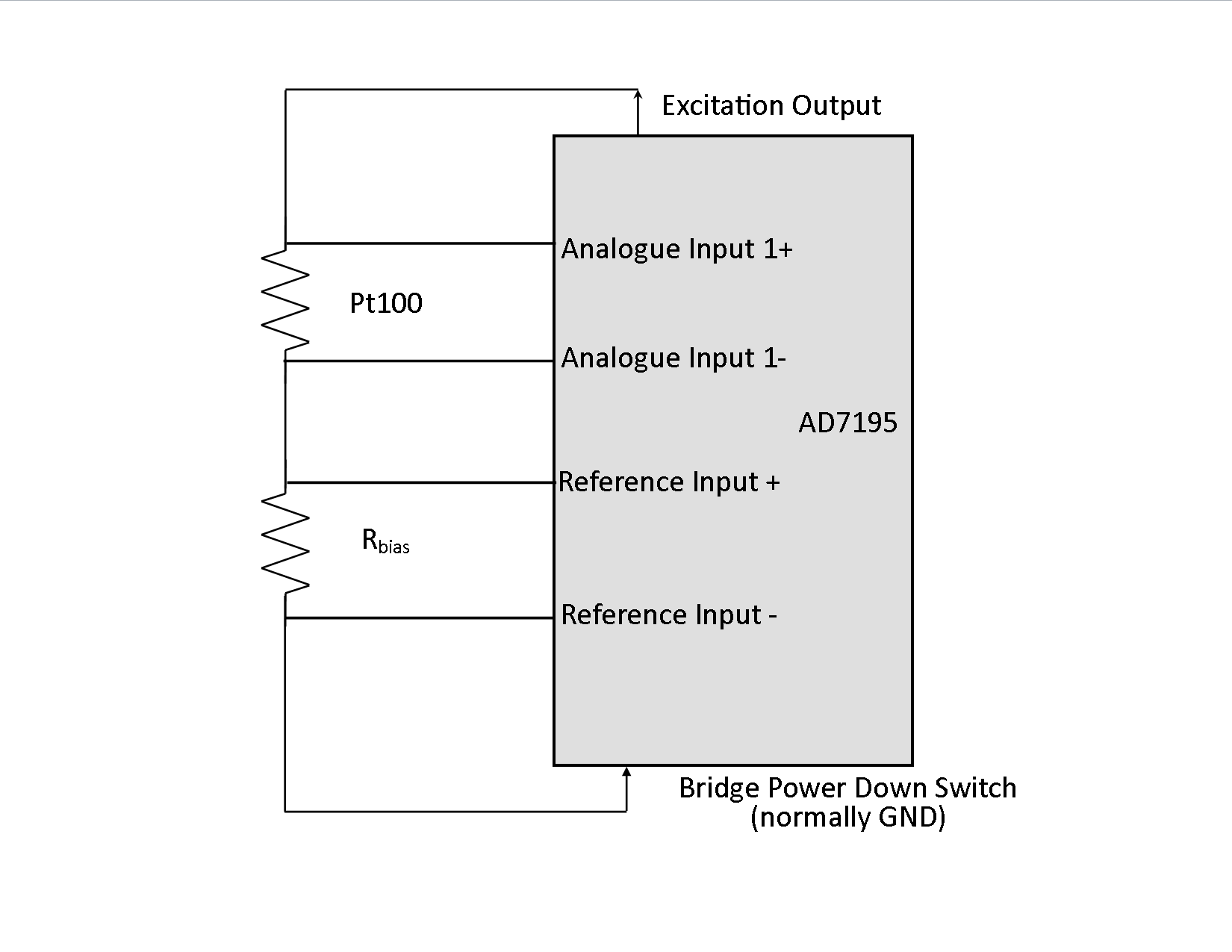}
	
	\caption{An example of a ratiometric measurement. A voltage is applied to one side of two resistors in series allowing the AD7195 to measure the voltage drop across each. Due to one of the resistors in the bridge being thermally stable this allows a calculation of the resistance to be measured. The output from the AD7195 would be the ratio of the two resistances, thus multiplying by the bias resistors resistance, a value of the PT100 is obtained allowing the temperature to be found using equation \protect\ref{eq-Temperature}.}
	\label{fig-ratiometric}	
	\end{figure}
\end{center}

One technique to measure resistance is the 4-wire measurement, which offers more advantages compared to the 2-wire measurement. The 2-wire measurement involves the measurement of a voltage over a resistor when applying a constant current, however, fluctuations in the current will cause changes in the measured voltage and appear to be a fluctuation in temperature. To combat this, a voltage can be applied to two resistors in series, one PT100 and one stable bias resistor. This stable bias resistor has a thermal coefficient of $50$ parts per billion (ppb), meaning a $10$ k$\Omega$ resistor would change by $0.5$ m$\Omega$ with a $1$ K change in temperature. When a measurement is made, it is always done as the ratio of the voltage drops over each resistor. This would cancel any current/voltage noise due to it being in common mode. This is known as a 4-wire measurement or ratiometric measurement. The simplified scheme is presented in figure \ref{fig-ratiometric}. It also has an added advantage that any resistances in the wires used are also cancelled to a large degree.

Here, an AD7195\cite{AnalogAD7195} was used due to its ability to not only automatically calculate the ratiometric measurement, but also to output an AC excitation and internally switch inputs so that valid data is converted. This AC excitation removes effects such as bridge potentials that occur when a DC current is used. These bridge potentials can offset the data, resulting in a misleading value of temperature, as well as drifts in the data. The result from the AD7195 which is the ratio of the PT100 and the bias resistor, is converted to a resistance of the PT100 by multiplying it by the resistance of the bias resistor, i.e. $10$ k$\Omega$. Since the resistance of the PT100, is around $100$ $\Omega$, which is a factor of $100$ lower than the bias resistor, this allowed a gain of $64$ to be utilised on the AD7195s to obtain an improved noise performance on the input signal.

\subsection{Temperature Measurement Noise}

There are several sources of error in measuring a temperature. Though the effect is small, the resistance of the bias resistor can still change due to temperature, which would be observed as a change in measured temperature. For a bias resistor with a thermal coefficient of $50$ ppb, this would mean a $1$ K change in temperature would also change the signal by $50$ ppb. This $50$ ppb would result in approximately a $5.4$ $\mu\Omega \equiv 14$ $\mu$K error in the PT100 measurement. This is negligible under most circumstances, even if the temperature changes by $20$ K, as other sources of noise exceed this value. The input of the AD7195 also introduces an error of $3$ bits peak to peak at a gain of $64$ and a sampling rate of $4.7$ Hz (as stated in page 14 of datasheet\cite{AnalogAD7195}). A $3$ bit error on the input is the equivalent of $8$ in decimal which is the same as $4.8$ m$\Omega$. $4.8$ m$\Omega$ is the equivalent of $12.4$ mK pre-filtered. Using the same simulation used for shot noise, an estimate of the shot noise after filtering can be made. Here, $12.4$ mK peak to peak input noise results in $\approx 1.4$ mK peak to peak after three stages of decimation. The estimate seems to be smaller than the measured value, likely due to the stated values in the data sheet not being exact for all chips. 

\subsection{Temperature Control}

\begin{center}
	\begin{figure}
	
		\includegraphics[width=0.45\textwidth]{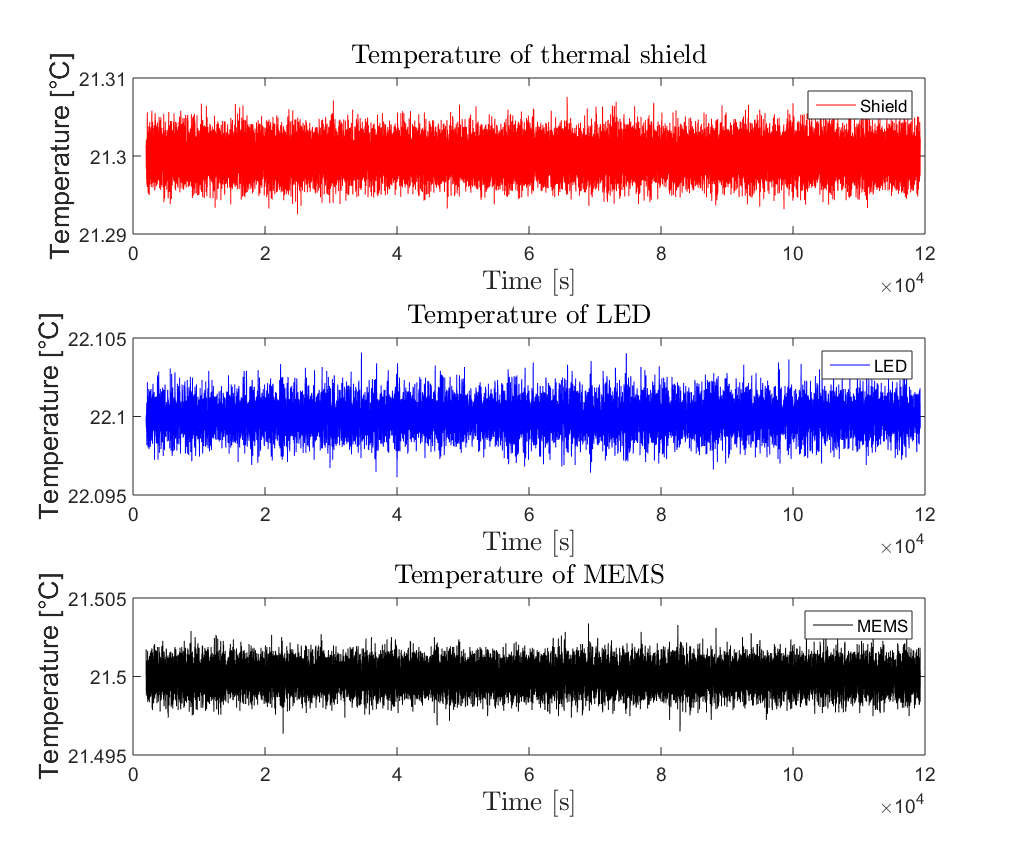}
		\caption{Three graphs showing the temperature stability of the Shield (Top), LED (Middle) and MEMS (Bottom) over $\approx 30.5$ hours. The largest variations are in the shield, which is the hardest to control, and are at $\approx \pm 5$ mK whereas the LED and MEMS vary by $\approx \pm 2$ mK}
		\label{fig-tempcontrol}
	\end{figure}
\end{center}

The values obtained from the temperatures are then filtered and decimated to obtain a data period of $\approx 1 \rightarrow 3$ s. These can then be used in a control system like the proportional, integral and differential (PID) that is currently used\cite{KJAFeedbackSystems}. This PID is external to the board, it is operated from the PC software with which the board communicates. This will later be added as a function onto the  firmware of the board. The PID control allows multiple temperatures to be simultaneously controlled (as seen in figure \ref{fig-tempcontrol}), where two temperatures, the MEMS, and LED are controlled to within $\pm 2$ mK, and the outer shield that helps thermally insulate the device to $\pm 5$ mK over a day. The shield is more difficult to control as its larger surface area can couple easier to the steel vacuum system (which is not thermally isolated) whereas the LED and MEMS are being thermally controlled inside the shield. 

\begin{center}
	\begin{figure}
	
		\includegraphics[width=0.45\textwidth]{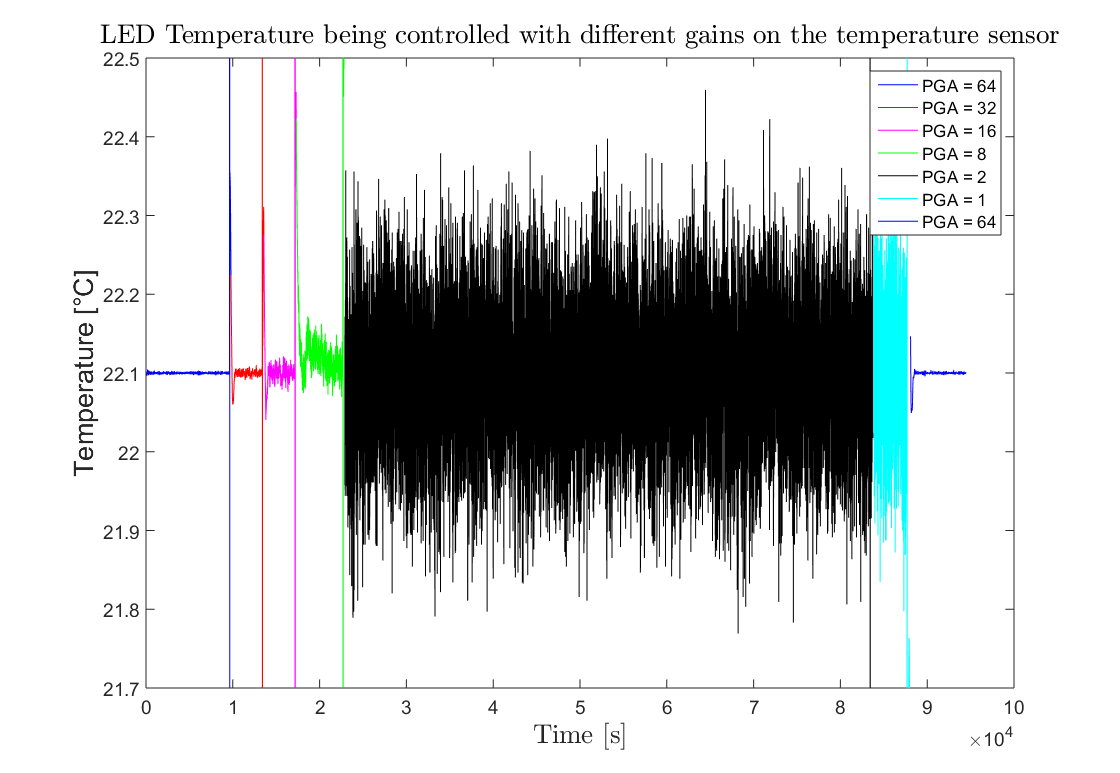}
		\caption{Comparison of the LED temperature when using different gains on the AD7195 input. It shows that decreasing the gain increases the noise on the control due to the system being limited by the input noise of the ADC.}
		\label{fig-tempcontrol_PGA}
	\end{figure}
\end{center}

It can be seen from figure \ref{fig-tempcontrol_PGA} that on decreasing the gain from the optimal value of $64$, increases the noise on the measurements. The changing of the gain implies that the system is limited purely by the input noise of the AD7195.

\section{Tilt Sensors}

As the device is measuring the component of gravity, any tilt off-axis of the proof mass relative to the direction of gravity will change the measured output of the device. This makes tilt an important variable to monitor. It also has a secondary effect in changing the resonant frequency of the MEMS proof mass that results in a different displacement for a given gravitational acceleration (seen in equation \ref{eq-tilt}).

\begin{equation} \label{eq-tilt}
x = \frac{g\cos{\theta}}{\omega^2}
\end{equation}

\noindent where $x$ is the displacement of the MEMS proof mass in m, $g$ is the gravitational acceleration felt by the MEMS proof mass in m s$^{-2}$, $\omega$ is the resonant angular frequency of the device in rad~s$^{-1}$ and $\theta$ is the angle relative to the vertical direction. By differentiating equation \ref{eq-tilt} with respect to changes in \textit{g}, $dg = \frac{-g\sin{\theta}}{\omega^2}$ is obtained. From this it would be assumed that for small $\theta$ the device is insensitive to changes in tilt, however, this is misleading and it is much more meaningful to consider $\cos{\theta_1} - \cos{\theta_2}$ instead of $\sin{\theta}$ when working with small changes in tilt as even if the device was off axis by $1$ $\mu$rad then the differential would not be significant.

\begin{center}
	\begin{figure}
		
		\includegraphics[width=0.45\textwidth]{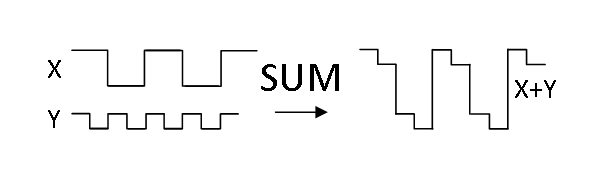}
		
		\caption{Figure shows the excitation used for each tilt axis, X and Y, and what shape the data takes when combining each of the components. X and Y are summed together and sampled by an ADC at each of the four phases and allows each of the axes to be extracted from a single input. The figure gives an example where the tilt is larger in X than Y, which, can also be extracted from the the right when comparing the amplitude of modulation of each of the frequency components.}
		\label{fig-TiltXYSUM}		
	\end{figure}
\end{center}

\begin{center}
	\begin{figure}
		
		\includegraphics[width=0.45\textwidth]{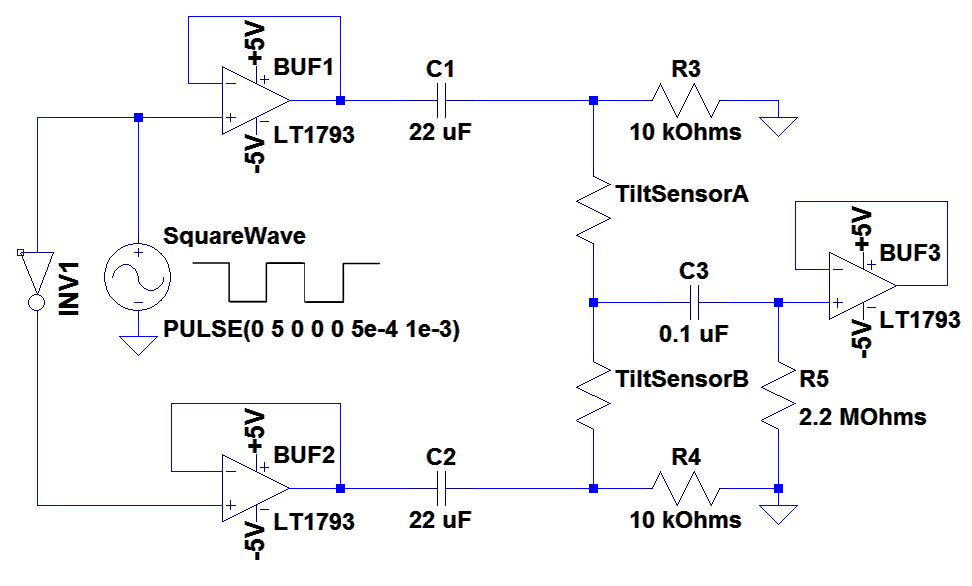}
		
		\caption{Schematic diagram for running the Spectron electrolytic tilt sensors (the SH50055-A-009). An anti-phase drive is produced from the micro-controller's digital outputs to produce a voltage difference over the tilt sensor inputs. The output sees the voltage over one of the resistors during one half of a cycle and the other during the second. Taking the difference of these voltages gives a value based on how far its been tilted, i.e. the resistance of each side changes with tilt. This circuit is replicated for the other axis of measurement but with a square wave of twice the frequency. These components are summed together and read by an internal ADC on the micro-controller. Each component can be extracted as information on which part of the cycle is known by the micro-controller.}
		\label{fig-TiltSensor}		
	\end{figure}
\end{center}

Using the outputs from the dsPIC itself to obtain high switching rates, an anti-phase oscillating signal is passed into two inputs of the electrolytic tilt sensor, SH50055-A-009\cite{SpectronSH50055}. This anti-phase configuration (presented in figure \ref{fig-TiltSensor}) gives an output that oscillates between the voltage over one side of the electrolytic tilt sensor, then the other, allowing a difference of the sides to be taken. This difference is calculated digitally by the micro-controller after it is digitised, using an ADC internal to the dsPIC to sample. The difference in the measured voltage between the measured voltage from being at the high state to the low state is a measurement of the difference in impedance of each side of the bridge. This difference in impedance relates to the angle of the sensor from being level, where zero difference between the high and low state would be measured when level. Figure \ref{fig-TiltSensor} demonstrates the complete circuit for one axis, another copy of this circuit is required for the other axis but is modulated with a square wave of twice the frequency. These two axis outputs are then summed together before being sampled by the  internal ADC of the dsPIC. Each axis can then be extracted from the single value by examining the phase each frequency component is at when sampling (seen in equation \ref{eq-tiltdemod}).

\begin{subequations} \label{eq-tiltdemod}
	\begin{align}
		V_{X} &= \frac{(V_1 + V_2) - (V_3 + V_4)}{2} \\
		V_{Y} &= \frac{(V_1 + V_3) - (V_2 + V_4)}{2} 
	\end{align}
\end{subequations}

\noindent where V$_X$ and V$_Y$ are the voltage relating to the tilt of each axes respectively and V$_n$ is the voltage measured by the ADC during a four phase output cycle. Here axis Y is sampled at twice the frequency of X and this is illustrated in figure \ref{fig-TiltXYSUM}. The figure shows a small tilt in the Y axis being summed onto a larger tilt in the X axis.

\begin{center}
	\begin{figure}
	
	\includegraphics[width=0.45\textwidth]{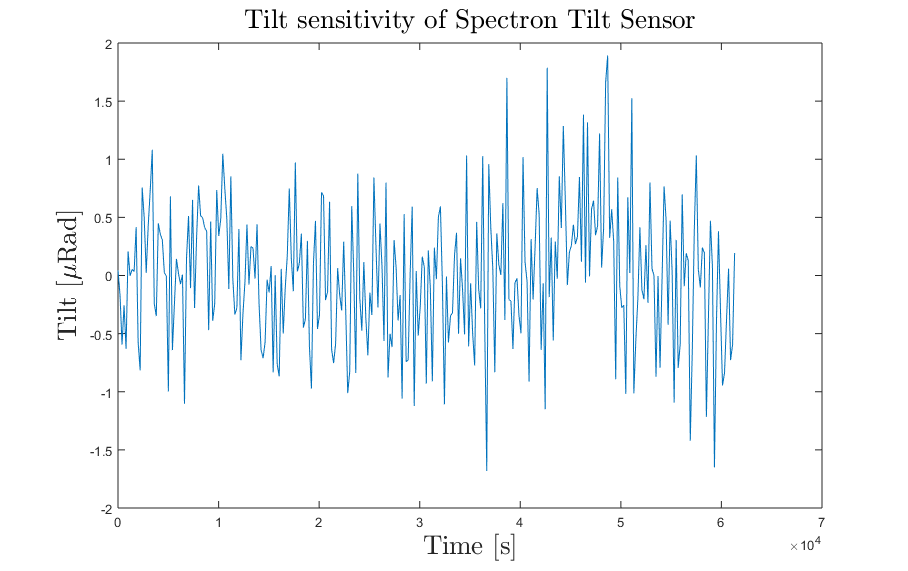}	
	
	\caption{Figure shows a maximum change of $\approx 2$ $\mu$rad over $\approx 16.7$ hours. This data has had low frequency polynomial drift removed using a regression.}
	\label{fig-tilt}
	\end{figure}
\end{center}

For every $1$ $\mu$rad; $2.2$ mV is obtained, i.e. $2.2$ V/mrad. This allows for a calibration of the tilt sensitivities by tilting it through a known amount and monitoring how much the output of the signal changes. The MEMS proof mass has a tilt sensitivity of $4.4$ $\mu$Gal/$\mu$rad. This requires the proof mass to have to tilted relative to the vertical of less than $\approx 9$ $\mu$rad to get the desired $40$ $\mu$Gal sensitivity. Figure \ref{fig-tilt} demonstrates a stability of $\pm 1$ $\mu$rad.

\section{Conclusion}

A highly stable optical shadow sensor has been demonstrated along with its digital readout and control. Whilst the shadow sensor has been presented to be used in a small, battery-powered, cost efficient and lightweight MEMS gravimeter, the shadow sensor could be re-purposed to serve in many precision sensing applications. The system is shown to be able to measure displacements with a sensitivity of $1 \rightarrow 2$ nm/$\sqrt{\textrm{Hz}}$ at $1000$ s over a period greater than a day. All functionality is obtained from a micro-controller based, custom electronics board that: measures and controls several temperatures to $\pm 2$ mK; monitors changes in tilt to $\pm 2$ $\mu$rad; modulates an LED using four points per cycle; converts $\mu$A of current to measurable voltages; demodulates a digitised signal (a micro-controller based digital lock-in amplifier); computes digital filters, and decimates measured variables to lower data rates with decreased noise. Further work is ongoing on a new electronics board to further improve the displacement sensitivity of the device and to reduce the power consumption. One such change is replacing the ADS1248, used to measure the signal, with a faster and simultaneous lower noise  sampling analogue to digital converter (AD7768). This could also allow testing of a system where the reference and modulation is generated off board, which requires very closely timed sampling of both the signal and reference for demodulation. The AD7195 temperature sensor is being changed to allow access to all the signal inputs. This will mean that more temperature measurements can be taken albeit at a lower sampling rate per channel.

%



\section*{Acknowledgment}

\ifCLASSOPTIONcaptionsoff
  \newpage
\fi



%

\bibliographystyle{IEEEtran}

\end{document}